\documentclass[12pt]{article}
\usepackage{a4}

\begin{document}

\begin{titlepage}

\begin{center}
{\Large\bf Neutrino Sector with Majorana Mass Terms and Friedberg-Lee
Symmetry}
\vfill

{\bf C. Jarlskog}\\[0.3cm]

%
%
{Division of Mathematical Physics\\
LTH, Lund University\\
Box 118, S-22100 Lund, Sweden}

\end{center}
\vfill
\begin{abstract}

We examine a recently proposed symmetry/condition 
by Friedberg and Lee 
in the framework where three right-handed
neutrinos are added to the spectrum of the three-family 
Minimal Standard Model. It is found that the right-handed
neutrinos are very special, with respect to this symmetry.
In the symmetry limit the neutrinos are massless and
that may be a hint about why they are light. Imposed as a condition
and not as a full symmetry, we find that one of 
the three right-handed neutrinos simply
decouples (has only gravitational
interactions) and that there is a massless
interacting neutrino. The possible relation
of the model to the see-saw mechanism is briefly discussed.

\end{abstract}
\vfill
{\bf PACS: Neutrinos, Masses, Mixings}
\vfill
\end{titlepage}

\section{Introduction}

The question of the origin 
of the quark and lepton masses is considered to be
one of the most important ones in physics. 
However, in spite of a huge amount of effort during 
the past three decades, we do not understand the pattern of
the fermion masses and mixings. This area of research is
badly in need of new ideas. 

Recently, Friedberg and Lee have in a series of papers
(\cite{FL1} - \cite{FL3})
addressed this question, by proposing a translational
symmetry for fermion fields (see below).  
We find that their proposal merits
further thought and, therefore, in this short paper we take
a closer look at it in the framework of the three-family
Standard Model with the addition of three right-handed neutrinos.
We extend the Friedberg-Lee analysis by adding Majorana mass
terms for the right-handed neutrinos. 

The plan of this article is as follows. First, we briefly recapitulate 
the Friedberg-Lee proposition  and then
consider some alternative approaches to implementing it,
in the framework of the Standard Model.
We find that the right-handed neutrino sector is very
special, due to the fact that these objects do not interact
with the gauge fields. Consequently, we examine the case of neutrinos
in more detail and present our conclusions in the last section.

\section{The Friedberg-Lee Proposition}

Friedberg and Lee take the fermion mass terms, in
the Lagrangian, to be of the form
\begin{equation}
L_{mass} = \overline{f_j}M_{jk}f_k
\end{equation}
where, for example, for the up-type quarks the $f_j, ~j=1-3$,
represent three fermion fields with the quantum numbers of the
up-type quarks and $M$ is the three-by-three Hermitian
mass matrix of these quarks.
This mass term is required to be invariant
under
\begin{equation}
f_j \rightarrow f_j + \eta_j \theta
\end{equation}
where $\eta_j$ are constants, to be determined, and $\theta$ is
an anticommuting (Grassmann) parameter, $\theta^2 =0$, which
is independent of space and time.

There are four different $L_{mass}$ terms in the Lagrangian, 
corresponding respectively to up-type quarks, down-type quarks,
charged leptons and neutrinos. Therefore, there are,
in the Friedberg-Lee scheme four different
$\theta$'s, as well as four sets of $\eta$'s.   

The Friedberg-Lee requirement yields that
$M_{jk} ~\eta_k =0$.
In the symmetry limit, i.e., when this relation is to hold
for arbitray $\eta$, the 
mass matrix would identically vanish. Therefore, the condition
is imposed for a set of $\eta$'s, to be determined.
This implies that $M$ has a zero eigenvalue, i.e., $m_u =0$,
for the case of up-type quarks. Indeed the up-quark
mass is much smaller than the charm and the top quark masses
and, therefore, one could imagine that it were actually zero,
in the lowest order in some more elaborate scheme. 
Similarly, the down quark, the electron and one of the neutrinos
are massless.
These results are again interesting as $m_e/m_{\mu}$
and $m_d/m_s$ are small numbers. 
With one massless fermion in each sector at the departure point,
Friedberg and Lee have invested substantial effort into 
the needed modifications for getting a realistic model. See 
references \cite{FL1} - \cite{FL3}. 

\section{Possible Approaches in the 
Standard Model}

In the Standard Model the situation looks somewhat more
complicated because the left-handed and right-handed
fermions have a priori nothing to do with one another.
One could say that they are born in different worlds
and meet each other through the Higgs field, 
the mass matrices being their coupling constants (up to an
overall factor) to the Higgs field. These matrices are
in general not Hermitian. 
We shall now consider how these matters may affect the
Friedberg-Lee conditions.  

Consider the Lagrangian of the Standard Model which
symbolically can be written in the form
\begin{equation}
L=L(F,G) + L(F,H) + V(H) +  L(G) + L(H,G)
\end{equation}
Here $F,~G$ and $H$ denote the fermion, gauge and 
Higgs fields respectively. $L(F,G)$ gives
the kinetic terms of the fermions and their interactions
with the gauge bosons and $L(F,H)$ contains the fermion
mass terms. The last three terms have been written down
for completeness but will not directly concern us here.  

The fermion mass terms, 
for quarks and charged leptons, are of the form
\begin{equation}
L^{SM}_{mass} = \overline{f_{jL}}M^{Dir}_{jk}f_{kR} + h.c.
\end{equation}
In the literature, these are called Dirac mass terms, being
the only possible terms for these 
charged particles.

For neutrinos the situation is more complicated. Introducing
right-handed neutrinos results not 
only in Dirac mass terms
but also in Majorana mass terms
\begin{equation}
L^{Maj} = \overline{f^C_{jR}} M^{Maj}_{jk}f_{kR} + h.c.
\end{equation}
where in this equation $f$ stands for the neutrino field.

The left-handed
and right-handed fermions being unrelated to each other implies 
that there are several options 
for implementing the Friedberg-Lee
proposition. One could introduce it for

1. only the left-handed fermions

2. only the right-handed fermions

3. both chiralities

One could also choose some mixtures of the above.
We find that the option 2 is the most attractive one.
The reason being that 
the right-handed fermions have less gauge interactions.
In option 1, the condition would forbid all interactions
of the left-handed quarks and/or leptons. 
The right-handed quarks and charged leptons
do not interact with the $W$'s. Thus in option
2 "only" their 
interactions with the $B$ field would be forbidden. 
The right-handed neutrinos are in this respect
very special as they have no 
interactions at all with the gauge fields.
They constitute a basis for maximal symmetry. 

For the case of quarks and charged leptons 
we have nothing to add to the 
Friedberg-Lee analysis as
the assumption that the above condition be valid
only for right-handed fermions (or for that matter only
for left-handed fermions) leads to 
a zero mass state in each of these sectors.
Therefore, from now on, we focus our attention on the 
neutrino sector.

\section{Friedberg-Lee Condition in $\nu_R$-sector}

Consider the transformation
\begin{equation}
\nu_{Rj} \rightarrow  \nu_{Rj} + \eta_j \theta
\end{equation}
Here $\eta_j, ~j=1-3$ are in general complex numbers, and
$\theta$ carries the quantum numbers of the right-handed 
neutrinos (weak isospin and hypercharge equal to zero).

A particularly attractive feature of this transformation
is that ${\underline{ all}}$ the other terms in 
the Standard Model Lagrangian are
invariant under it, with arbitrary $\eta$. 
More precisely, the kinetic terms for the right-handed 
neutrinos are not invariant but since $\theta$ does not
depend on space and time the resulting action is invariant.
Thus the above translation is a symmetry of the Standard 
Model excluding the neutrino-Higgs terms in $L(F , H)$
and the Majorana mass terms.
Therefore, requiring  
the above translations to be a symmetry of the entire action  
immediately yields that the neutrino masses
are all zero. This is a welcome result as it perhaps
may explains why that neutrino masses are so small. They
start off by being zero in the symmetry limit. One could
imagine that there is a scenario beyond the Standard Model in
which the above symmetry holds and somehow in the broken
version the $\eta$'s get fixed. 

In the real world the $\eta$'s are not arbitrary as the
neutrinos are not massless. Let us consider, in somewhat
more detail, what happens
in the Standard Model, with three right-handed
neutrinos. The neutrino mass matrix is then of the form 
\begin{equation}
L^{(\nu)}_{mass}= -{1 \over 2} (\overline{\nu^{}_L}, \overline {\nu^C_R}) 
{\cal M} \left( \begin{array}{c}
\nu_L^C \\ \nu_R \end{array} \right) + h.c.
\end{equation}
where
\begin{equation} \nu_X = \left(
\begin{array}{c} \nu_{1X} \\ \nu_{2X} \\ \nu_{3X}
\end{array} \right) 
\end{equation}
$X=L, R$. Here
${\cal M}$ is the six-by-six neutrino mass matrix
given by
\begin{equation} 
{\cal M}  = \left(
\begin{array}{cc}
0 & A  \\
A^T & M 
\end{array}
\right)
\end{equation}
${\cal M}$ is a 
symmetric matrix; 
$A$ and $M$ being respectively the three-by-three 
Dirac and Majorana mass matrices and $T$ stands for transpose. 
For simplicity we have
suppressed the superscripts $Dir$ and $Maj$. 

The Friedberg-Lee condition that
$\nu_{Rj} \rightarrow  \nu_{Rj} + \eta_j \theta$ 
should leave the Lagrangian invariant implies
\begin{equation}
A_{jk} \eta_k =0, ~~~
M_{jk} \eta_k =0
\end{equation}
We may now redefine the right-handed neutrino fields
by
\begin {equation}
\nu_R \rightarrow U \nu_R
\end{equation}
where $U$ is a three-by-three unitary matrix.
Under this transformation the Lagrangian retains it's
form. All that happens is a redefiniton of the mass matrices 
\begin{equation}
M \rightarrow M^\prime= U^{\star} M U^\dagger, ~~~
A \rightarrow A^\prime= A U^\dagger
\end{equation}
The Friedberg-Lee condition now reads
\begin{equation}
A^\prime_{jk} \eta^\prime_k =0, ~~~
M^\prime_{jk} \eta^\prime_k =0
\end{equation}
where $\eta^\prime = U \eta$. 
We may choose $U$ such that $M^\prime$ is diagonal. 
Since $M^\prime$ has a zero
eigenvalue, by permutation of the right-handed neutrino
fields (which leaves the rest of Lagrangian invariant) 
it may be written in the form
\begin{equation}
M^\prime = \left( \begin{array}{ccc}
M_1 &0 & 0\\ 
0 & M_2 & 0 \\
0 & 0 & 0
\end{array} \right) 
\end{equation}
Assuming that $M_1$ and $M_2$ are nonzero, we find
that the vector $\eta^\prime$ is proportional to $(0, 0, 1)$.
This in turn implies that the third column in matrix 
$A^\prime$ is zero. Therefore it is of the form
\begin{equation}
A^\prime = \left( \begin{array}{ccc}
a_{11} &a_{12} & 0\\ 
a_{21} & a_{22} & 0 \\
a_{31} & a_{32} & 0
\end{array}
\right)
\end{equation}
A remarkable consequence of the Friedberg-Lee condition
is that one of the three right-handed neutrinos simply
decouples, in the sense that it has only gravitational
interactions due to its kinetic energy term.
We end up with three left-handed neutrinos
and just two right-handed neutrinos. The mass matrix is then of
the form
\begin{equation}
{\cal M} = \left( \begin{array}{ccccc}
0 & 0 & 0 & a_{11} &a_{12} \\ 
0 & 0 & 0 & a_{21} & a_{22}  \\
0 & 0 & 0 & a_{31} & a_{32} \\
a_{11} & a_{21} & a_{31} & M_1 & 0 \\
a_{12} & a_{22} & a_{32} & 0  & M_2
\end{array}
\right)
\end{equation}
To sum up, in this model there is, in
addition to the non-interacting massless neutrino,
a massless interacting neutrino. The masses and
mixings will depend on the choice of the parameters.
A natural scenario 
would be to have small values for
magnitudes of $M_1$ and $M_2$ because in the symmetry limit
these are zero. In the limit $M_1, M_2 \rightarrow 0$
one obtains three Dirac neutrinos, one of them being massless.
The model may or may not violate CP symmetry depending
on how the charged lepton mass matrix looks like.
The if and only if condition for CP violation in this
case is as for the quarks, that the determinant of the commutator
of the respective mass matrices be nonzero \cite{ceja85}. 

However, since we don't really know what
is natural or not, 
the model also allows see-saw mechanism, which is obtained 
by taking the magnitudes of $M_1$ and $M_2$ to be very
large. Then there will be two very heavy neutrinos
that do not interact with the gauge bosons
and three light ones, one of them being 
massless. The phenomenology of  
see-saw models with two right-handed neutrinos
has been the subject of several studies in the literature
(see, for example \cite{ross} and \cite{ibarra}
and references cited therein).

It should be mentioned that the fermionic translations
may look "innocent" but are far more subtle than their
bosonic counterparts, such as 
$\phi \rightarrow \phi + v$, or $A_\mu \rightarrow
A_\mu + \partial_\mu \Lambda$. For example, under 
Lorentz transformations
$\theta$ must transform as a fermionic field, i.e.,
$\nu_R \rightarrow S ~\nu_R$ and $\theta \rightarrow S
~ \theta$, where $S$ is the appropriate transformation
matrix. An important point to keep in mind is that  
a nonzero vacuum expectation value of a fermionic operator
will break Lorentz invariance. One could imagine
that the Friedberg-Lee translation is actually driven
by the requirement of Lorentz invariance. However, this
matter requires further study which is beyond the scope of
the current paper. We would only like to add that possible mechanisms
for the violation of Lorentz invariance in particle
physics have been studied by many authors. For example,
Coleman and Glashow \cite{coleglas} have suggested 
adding tiny renormalizable terms, 
that violate Lorentz invariance, into the Lagrangian of
the Standard Model and have examined their phenomenological 
consequences for many processes, among them neutrino oscillations.
The work of these authors has led to a wave of further studies. Currently,
violation of Lorentz invariance is a rather 
popular domain of research.

\section{Conclusions and Outlook}

In this paper, we have examined the Friedberg-Lee proposition,
that the mass terms in the Lagrangian be invariant under a prescribed
translation of fermion fields. 
For the case of quarks and charged leptons, it does not matter whether
the constraint is imposed on left-handed or the right-handed fermions.
The result, in each case, is as obtained by Friedberg and Lee. In the
neutrino sector, we find that the right-handed neutrinos are very special
because the action is invariant under the translation,
$\nu_{Rj} \rightarrow  \nu_{Rj}+ \eta_j \theta$, with the exception 
of those terms which involve neutrino masses and interactions
with the Higgs particle. Adding three right-handed neutrinos, to 
the three-family version of the Minimal Standard Model and taking into
account both the Dirac and Majorana mass terms, 
we find a very interesting result: the
Friedberg-Lee condition forces one right-handed neutrino to decouple.
The ensuing model is equivalent to introducing only two right-handed 
neutrinos. 
See-saw-like model is allowed, with two heavy 
neutrinos and three light neutrinos, one being massless. 
But the Majorana mass
terms could also be very small. In this case, one could argue
that the neutrino masses
are small because in the symmetry limit (i.e., with $\eta$ being arbitrary)
the Friedberg-Lee condition requires neutrinos to be massless. 
One may wonder what happens 
with the neutrino counting, that gives the result $N_\nu$, at the Z-peak.
Doesn't one then get too large a value for this number? 
The current experimental value is
$N_\nu = 2.9840 \pm 0.0082$ \cite{lepslc}.
The answer is no. 
It has been shown (\cite{ceja90}) that adding any number of 
right-handed neutrinos
to the spectrum of the three family Minimal Standard Model
yields $N_\nu \le 3 $.

\end{document}